\documentclass[
     ,draft            
     ,numberedheadings 
  ]
  {aipproc}

\layoutstyle{6x9}
\def\lessim{\lower.5ex\hbox{$\; \buildrel < \over \sim \;$}}
\def\gtrsim{\lower.5ex\hbox{$\; \buildrel > \over \sim \;$}}

\begin{document}

\title{Hadronization and Quark Probes of Deconfinement at RHIC}

\classification{25.75.Nq,25.75.-q,24.10.Pa,25.75.Dw}
\keywords      {Quark Gluon Plasma,  Deconfinement, Relativistic heavy-ion collisions, Strangeness, Hadronization,  Collective Flow, Elliptic Flow, Charm production}

\author{Huan Z. Huang}{
  address={Department of Physics and Astronomy, University of California, Los Angeles, CA 90095-1547}
}

\author{Johann Rafelski}{
  address={Department of Physics, University of Arizona, Tucson, AZ 85721-0081}
}

\begin{abstract}\footnote{This report constitutes the  combined contribution of the authors
 to the "VI Quark Confinement
and the Hadron Spectrum" conference, held 21--25 September 2004 in Sardinia, Italy
covering their oral presentations and the `Deconfinement' discussion session; proceedings to be published by the American Institute of Physics.}
We discuss  experimental features 
of identified particle production from nucleus-nucleus collisions. These features 
reflect hadronization from a deconfined partonic matter whose particle 
formation scheme is distinctly different
from fragmentation phenomenology in elementary collisions. 
Multi-parton dynamics, such as quark coalescences or 
recombinations, appear to be essential to explain the experimental 
measurements at the intermediate transverse 
momentum of 2--5 GeV/c. Constituent quarks seem to 
be the dominant degrees of freedom at hadronization. 
 Heavy quark production  should help quantify deconfined matter properties. 

\end{abstract}

\maketitle

\section{Introduction}

The advent of the Relativistic Heavy Ion Collider (RHIC) at Brookhaven
National Laboratory (BNL) has turned the next page in search  for, and study of, the
Quark Gluon Plasma (QGP). Most recent dAu baseline reaction 
measurements by all four RHIC experiments~\cite{dAu-star,dAu-phenix,dAu-phobos,dAu-bhrams}
confirmed that a dense strongly interacting medium has been created in central AuAu
collisions at RHIC.  The QCD nature of the dense
matter created at RHIC and whether the current experimental evidence proves
the discovery of the QGP have been since under debate within the heavy ion physics
community~\cite{RBRC-1,RBRC-2,RBRC-3,RBRC-4}. We will address here primarily  the 
physics `soft' hadron  experimental results and the related  evidence for deconfinement.
 The major topics we address are:  
 1) features of hadronization and other evidence for a color
deconfined bulk partonic matter; 2) the QCD properties of the matter at
hadronization; and
3)charm  production and related future experimental measurements capable to
 further quantify the  properties of deconfined phase. 

Among topics we discuss in depth  are: mechanisms of hadronization,
 experimental status of strange
particle production  enhancement,  azimuthal parton flow anisotropy, bulk 
properties of dense hadron matter at parton fireball breakup, charm experimental status
at RHIC.

\section{Features of Hadronization Observed at RHIC}
\subsection{What we know about hadronization}
We observe, in the final state,
 a multitude of hadrons, irrespective of what happened and which
reaction system is observed. The paradigm emerged that
QCD color charges are confined and hadrons exist in color singlet state. Individual
reactions occur between leptons and/or quarks.  However, the conversion, {\it i.e.},
hadronization of quarks and gluons (partons) has not been understood based
on first principles. There are several  widely studied models:
\begin{enumerate}
\item
Hadron formation in elementary 
$e^{+}e^{-}$ and $qq$ (that is nucleon--nucleon) collisions has been described 
in the pQCD domain  in terms of several components. The particle
production process is 
factorized into parton distribution functions, parton interaction processes
and in final step, fragmentation functions for hadron formation.\\[-0.4cm]

 The fragmentation function
is assumed to be universal and can be obtained phenomenologically 
from $e^{+}e^{-}$ collision measurments.
A typical Feynmen-Field~\cite{FF} fragmentation process involves a leading parton
of momentum $p$, which fragments into a hadron of momentum $p_{h}$ whose
properties are mostly determined by the leading parton. The fragmentation
function is a function of variable $z=p_{h}/p,\ z\in (O,1)$\,. 
Note that 
baryon production is found to be significantly suppressed compared to the production of
light mesons: the baryon to pion ratio increases with $z$, but never
exceeds $20\%$~\cite{SLD}, see also the  $pp$-STAR results  in 
figure\,\ref{ratio} below.\\[-0.3cm]
\item
In the soft (low $p_\bot$) particle production region, the pQCD framework and
factorization break down, particles are believed not to be from
fragmentation of partons. In the elementary interactions physics,
string fragmentation models were inspired by the
QCD description of the quark and anti-quark interaction. 
The Lund string model is one of the popular hadron formation
models which has been successfully implemented in Monte Carlo description of
$e^{+}e^{-}$, $pp$, and nuclear collisions~\cite{LUND}. 
Note that, in the string fragmentation models, the baryon production
is also suppressed because baryon formation requires the clustering of
three quarks~\cite{lund-b}.\\[-0.3cm]
\item Soft multi particle production in particular in the $AA$ (nuclear) collision domain
is described within the (Fermi) Statistical Hadronization (SH) model. SH is   a model of 
particle production in which the birth process of each particle  
fully saturates (maximizes) the quantum mechanical probability amplitude, and
thus, all hadron yields are determined by the appropriate integrals of the
accessible phase space.\\[-0.4cm]

The  statistical hadronization  model   introduced in 1950~\cite{Fer50,Pom51}  
 has matured today to a full
fledged tool  for soft strongly interacting particle 
production,  capable to describe  in detail hadron  abundances once
 the mass spectrum of hadron resonances is included  \cite{Hag65}.
The key SH  parameters within the grand canonical formulation of particle phase space
are the temperature $T$ and, at a finite baryon density present in a $AA$ reaction
system,  the baryochemical potential $\mu_{\rm B}$. 
It is generally accepted that  as the energy of the colliding 
nuclei varies,   a wide domain  of $T$ and  $\mu_{\rm B}$ is explored, 
see figure\,2~in~\cite{Hag80}. Note that the ratio of baryons to mesons
is,  in chemical equilibrium, also   suppressed, 
since $e^{m_{\pi,K}-m_{\rm baryon}\over T}\ll 1$ \\[-0.3cm]
\item
The reaction picture of a soup of quarks in a dense expanding fireball inevitably 
triggers development of recombination and coalescence  models \cite{KMR86,RafDan87}. 
Recent experimental RHIC results triggered further developments  
on quark coalescence~\cite{voloshin,lin} and
recombination~\cite{Duke,Hwa} which all have the essence of multi-parton dynamics for
hadron formation, despite significant differences in details.

\end{enumerate}

\subsection{(Anti)Baryon yield}

\begin{figure}[htb]
   \centering
 \includegraphics[height=0.29\textheight,angle=-90]{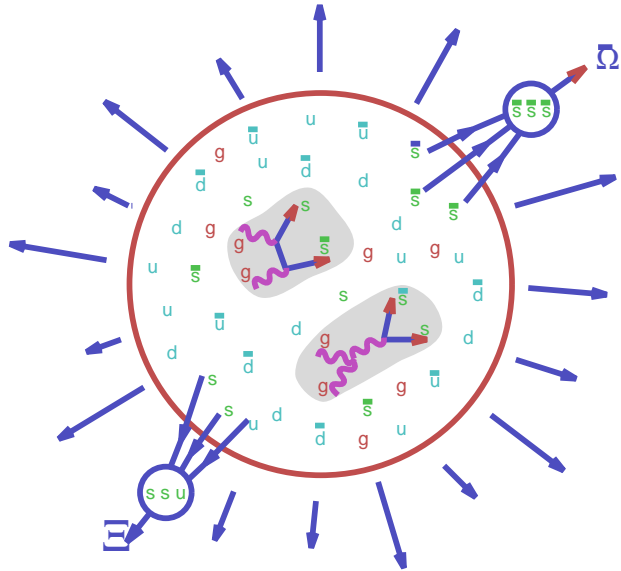} 
\caption{Illustration of the  two step mechanism of strange hadron formation 
from QGP: inserts show gluon fusion into
 strangeness, followed by QGP recombinant hadronization. }
\label{JRSTRANGEPROD}
\end{figure}

Strangeness plays a particularly
important role as a characteristic QGP observable.
The enhanced production of (strange) (anti)baryons 
has been an expected feature in recombinant hadronization
of QGP.  In nuclear collisions, the strange hadron yield derives
 from   two independent reaction steps following each other in time:
\begin{enumerate}
\item
the establishment of a ready supply of strange quark pairs  $s,\,\bar s$ 
which occurs predominantly in the initial  hot phase of  bulk partonic matter  
by the {\it thermal} pQCD gluon fusion processes  $gg\to s\bar s$   \cite{RM82}, 
in a manner  independent of the production of final
state hadrons;
\item  the high initial $s,\,\bar s$   yield survives the process of fireball
expansion evolution \cite{KMR86},  and 
contributes in  hadronization of pre-formed  $s,\,\bar s$ quarks
 to an unusually high multistrange (anti)baryon abundance~\cite{Raf82}.
\end{enumerate}
This is illustrated in figure\ref{JRSTRANGEPROD}.

Experiments indeed show very  significant new features of hadron formation
in $AA$ interactions compared to elementary 
$e^{+}e^{-}$ and $qq$ (that is $pp$) collisions. 
\begin{itemize}
\item 
Baryon production from $AA$ collisions, especially for
multi-strange hyperons, has been measured to be much larger than theoretical
(superposition, cascade) model calculations. 
The hyperon production per number of participant pairs
from $AA$ collisions at the SPS is significantly enhanced in
comparison with the value from $pp$ collisions. The enhancement factor
increases from $\Lambda $ to $\Omega $ hyperons and with the collision
centrality~\cite{WA97enh,Elia:2004mb}.
\item 
The  increase in the baryon production from  $AA$ collisions has become
much more prominent at RHIC energies. Figure~\ref{ratio} shows the ratios of
 $\overline{p}/\pi $ and $\overline{\Lambda }/K_{S}$ from central AuAu collisions at 
$\sqrt{s_{NN}}=130,$ and 200 GeV measured by PHENIX~\cite{phenix-b1} 
and STAR~\cite{star-b1,star-b2,star-k}. 
\begin{figure}[hbt]
 \includegraphics[height=0.35\textheight]{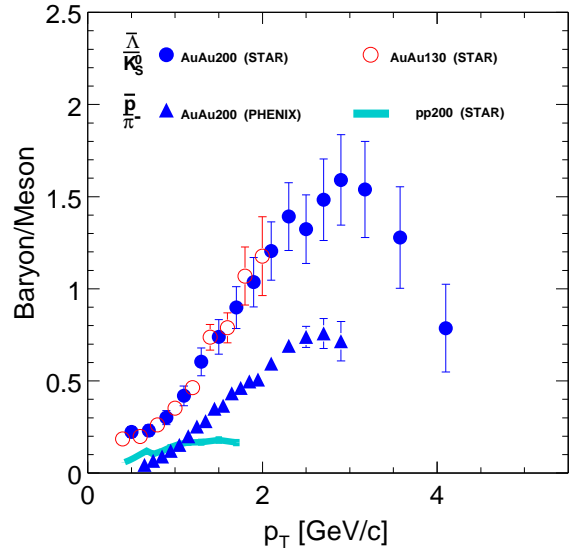} 
\caption{Ratios of ${\overline{\Lambda}}$ to $K_{S}$ from AuAu and $pp$ collisions (STAR) and 
$\overline{p}$ to $\pi$ from AuAu collisions (PHENIX) as a function of transverse momentum ($p_\bot$).
In addition to resonance contributions in all hadrons, the $\overline{\Lambda}$  includes 
contributions from $\Sigma^{0}$ decays.}
\label{ratio}
\end{figure}
\end{itemize}

The apparent difference,  in figure~\ref{ratio},
between the STAR and PHENIX ratios can be understood since   the
STAR $\overline{\Lambda }$ data include the electromagnetic decay contribution
from $\Sigma^{0}$ and both comprise different `towers' of hadron resonance decays. 
Note also that  the PHENIX data are corrected for post-reaction weak decay
contributions $Y,\overline Y\to N\overline N$. Similarly,
the STAR $\overline{\Lambda}$ data have been corrected 
for feed-down contributions from multi-strange hyperon decays.
Some early data, for example~\cite{phenix-b2}, are not shown because these 
data were not corrected for weak decay contributions.

The large baryon to meson ratio, seen in figure~\ref{ratio},  cannot be accommodated by the
traditional, {\it e.g.}, fragmentation scheme. 
The large ratio at the intermediate $p_\bot$ region provide 
clear evidence that particle formation dynamics in $AA$
collisions at RHIC are distinctly different from the traditional hadron formation
mechanism via fragmentation processes developed for the   elementary $e^{+}e^{-}$ and
nucleon-nucleon collisions.

The recombination models~\cite{Duke,Hwa} provided a satisfactory description of
the particle yields, in particular, the large production of baryons in the
intermediate $p_\bot$ region. The formation of a dense partonic system
provides a parton density dependent increase in baryon yield as a function of collision
centrality through the coalescence mechanism.

\begin{figure}[hbt]
\includegraphics[height=0.37\textheight]{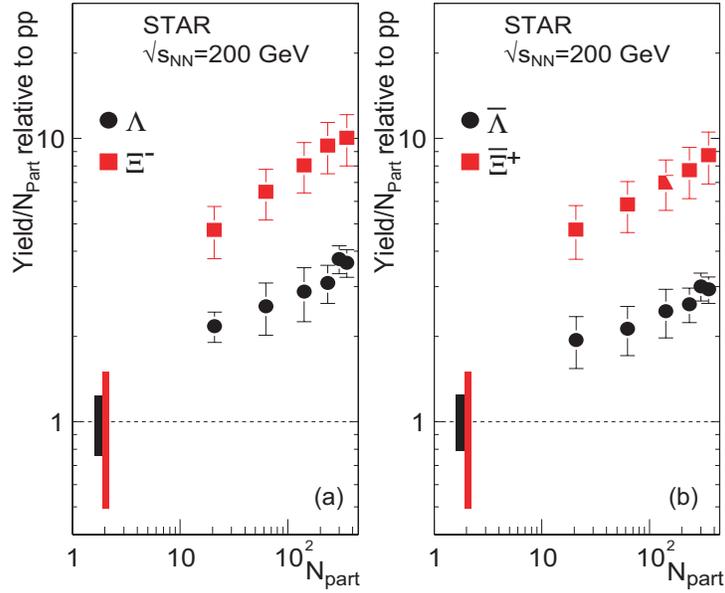} 
\caption{The STAR yields per participant  for production of 
 $\Lambda$ and $\Xi^{-}$ on left (a) and  $\bar{\Lambda}$ and $\bar{\Xi}^{+}$
on right (b) in AuAu collisions at $\sqrt{s_{\rm NN}}=200$\,GeV. 
Error bars are statistical.
Ranges for $pp$ reference data indicate the systematic
uncertainty.}
\label{RHICEnhance}
\end{figure}

Comparing  AuAu with $pp$   reactions at 
$\sqrt{s_{\rm NN}}=200$\,GeV,  see figure\ref {RHICEnhance},  we see
another large baryon production  enhancement  for strange
(anti)baryons $\Lambda, \overline\Lambda, \Xi^-,\overline{\Xi}^+$ 
reported by STAR \cite{CainCT}.
This enhancement  increases with centrality~\cite{Raf99,Let96}  and with greater
strangeness content  as found in strangeness recombination model ~\cite{KMR86}. 
These feature were recognized   early on as 
a characteristic signature of QGP \cite{Raf82}.

  This pattern of 
baryon enhancement has  been observed by the WA97 \cite{WA97enh} 
and NA57 \cite{Elia:2004mb} experiments at lower reaction energy 
available at the CERN-SPS. There is a gradual increase in
the strange antibaryon yield with reaction  energy \cite{Elia:2004mb}.
The large difference between baryon and antibaryon
 yields at SPS energy range are due to the presence of a
significant baryon density at lower reaction energy. For purpose of
comparison with RHIC results, this baryon density  effect can be removed by
 considering the geometric mean of the baryon and 
antibaryon yield.

In our opinion, the results presented above for the
systematics of strange hadron enhancement demonstrate 
that same novel mechanism operates in the entire collision
energy interval spanned by these data. Only a deconfined
quark-gluon plasma is a copious source of these often rarely 
produced hadrons. When the plasma fireball breaks up into final 
hadrons, the high abundance of  strange quarks and antiquarks  
manifests itself yielding high abundances of multi strange hadrons. 
 In conventional reaction schemes,  the production of particles containing two or 
more strange quarks is suppressed by the rarity of the required reaction processes.

\subsection{Nuclear modification factor}
  The nuclear modification factor
is defined in a collision number scaled  comparison of  peripheral with
 central head-on collisions:
\begin{equation}
R_{CP}=\frac{[yield/N_{bin}]^{central}}{[yield/N_{bin}]^{peripheral}}.
\end{equation}
 The nuclear modification factor
has also been defined by,
\begin{equation}
R_{AA}=\frac{[yield]^{AA}}{N_{bin}\times \lbrack yield]^{pp}},
\end{equation}
where $N_{bin}$ is the number of binary nucleon--nucleon collisions. The 
$[yield]^{AA}$ and $[yield]^{pp}$ are particle yields ($d^{2}n/dp_\bot dy$)
from AA and $pp$ collisions, respectively.

A $R_{AA}$ or $R_{CP}$ of unity implies that particle production from
$AA$ collision is equivalent to a superposition of independent
nucleon--nucleon collisions. Hard scattering processes within the pQCD
framework are believed to follow approximately the binary scaling in the
kinematic region where the nuclear shadowing of parton distribution function
and the Cronin effect are not significant. 

Measurements of charged hadrons and neutral pions have
revealed a strong suppression at high $p_\bot$ region in central
AuAu collisions~\cite{highpt-star1,highpt-star2,highpt-phenix1,highpt-phenix2}. 
Recent dAu measurements~\cite{dAu-star,dAu-phenix,dAu-phobos,dAu-bhrams} 
have demonstrated that the large
suppression of high $p_\bot$ particles in central AuAu collisions is mainly
due to energy loss, presumably of partons traversing the dense matter created
in these collisions.
\begin{figure}[htb]
 \includegraphics[height=0.30\textheight]{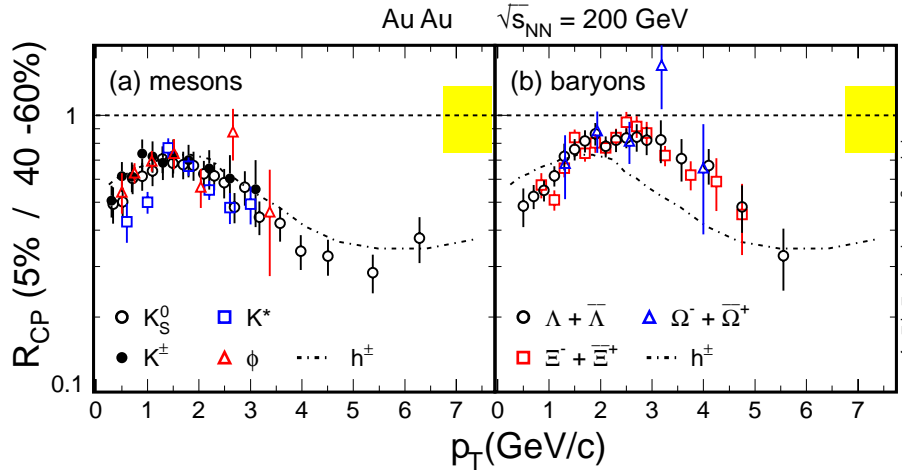} 
\caption{$R_{CP}$ of $K^{\pm}$, $K_{S}$, $K^{*}$, $\phi$,
$\Lambda$, $\Xi$ and $\Omega$ in comparison with charged hadron in dashed line. Distinct 
meson and baryon groups are observed.}
\label{fig:rcp}
\end{figure}

Figure~\ref{fig:rcp} shows the $R_{CP}$ of $K^{\pm}$, $K_{S}$, $K^{*}$, $\phi$, 
$\Lambda $, $\Xi $ and $\Omega $ as a function of $p_\bot$ from AuAu collisions at 
$\sqrt{s_{NN}}=200$ GeV measured by the STAR collaboration, where the ratio
is derived from the most central $5\%$ to the peripheral $40--60\%$ collisions. 
The dashed line is the 
$R_{CP}$ of charged hadrons for reference. In the low $p_\bot$ region, soft
particle production is dominated by the number of participant scaling.

The particle type dependence can be described using hydrodynamic flow
which predicts a mass dependence for the low $p_\bot$ spectra. In the intermediate 
$p_\bot$ region of 2 to 5.5 GeV/c the $p_\bot$ dependence of $R_{CP}$ falls
into two groups, one for mesons and one for baryons. Despite the large mass
differences between $K_{S}$ and $K^{\ast }/\phi $, and between $\Lambda $ and 
$\Xi $ little difference among the mesons and among the baryons has been
observed within statistical errors. 
Particle dependence in the nuclear modification factor disappears only above a
$p_\bot$ of 6 GeV/c,
consistent with the expectation from conventional fragmentation processes. The
unique meson and baryon dependence in the intermediate $p_\bot$ region
indicates the onset of a production dynamics very different from both
fragmentation at high $p_\bot$ and hydrodynamic behavior at low 
$p_\bot$.

\subsection{Azimuthal anisotropy}
\begin{figure}[htb]
\includegraphics[height=0.20\textheight]{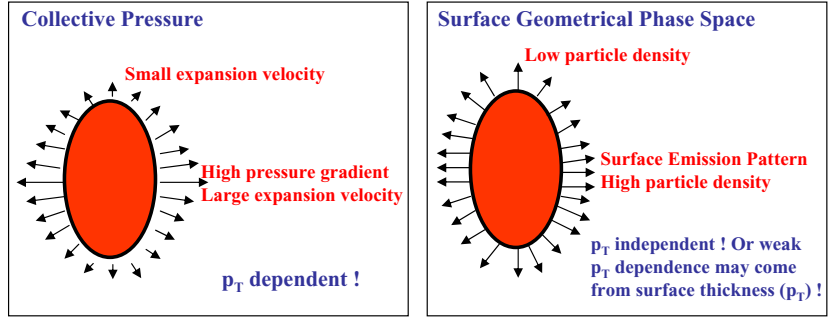} 
\caption{Schematic diagram to show two dynamical origins of angular anisotropy, one from
hydrodynamical expansion and the other from surface geometrical phase space.}
\label{schem}
\end{figure}

The azimuthal angular particle distribution can be described by a Fourier expansion, 
\begin{equation}
\frac{d^{2}n}{p_\bot dp_\bot d\phi }\propto (1+2\sum_{n}v_{n}\cos n(\phi -\Psi _{R})),
\end{equation}
where $\Psi _{R}$ is the reaction plane angle and $v_{2}$  has been called 
elliptic flow~\cite{Sorge}. 
In a non-central $AA$ collision the
overlapping participants form an almond shaped particle emission source, see figure~\ref{schem}.
The reaction plane is defined by the vectors $x$ (impact parameter direction) 
and $z$ (beam direction). The pressure gradient is greater and particles would experience 
larger   expansion force along the short axis direction (in plane) than  
along the long axis (out plane), see  figure~\ref{schem} left, resulting  in the final state 
in an ellipsoid in transverse momentum space. Theoretical, (typically hydrodynamic) model
calculations find  that this $p_\bot$ dependent component in
$v_{2}$ is generated mostly at the early stage of the nuclear collision.

There is another dynamical mechanisms responsible for $v_{2}$ due to  
the spatial geometrical component in the phase space of emitting source. 
  Particle production from a dense matter can be  squeezed more in the
reaction plane than that out of the reaction plane. More generally, 
a  freeze-out geometry will impose itself on particle abundance ellipticity. However,
any $p_\bot$ dependence would be a 2nd order effect associated {\it e.g.} 
with the optical depth of the freeze-out system. 
 Figure~\ref{schem} shows a
schematic diagram for these two mechanisms  of generating azimuthal
angular anisotropy $v_{2}$ in non-central collisions.

\begin{figure}[htb]
\includegraphics[height=0.35\textheight]{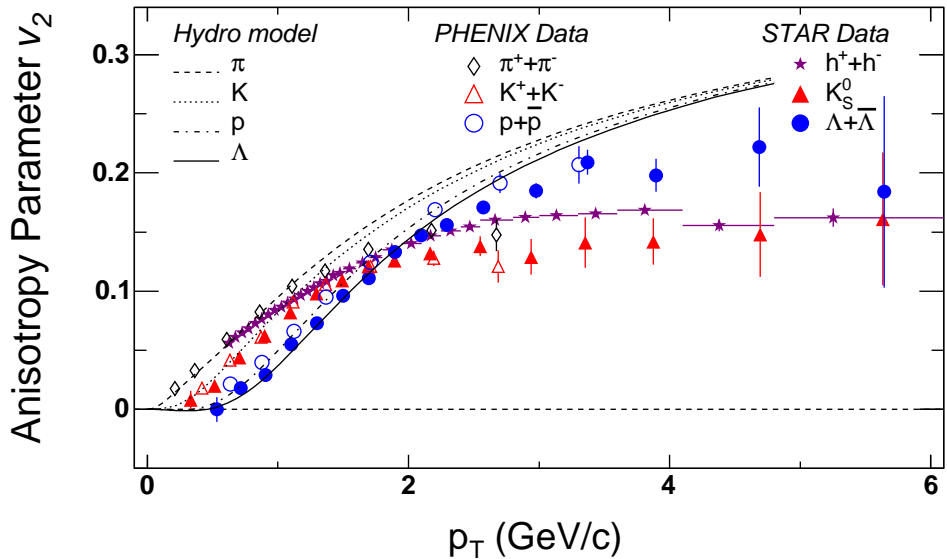} 
\caption{Azimuthal angular anisotropy $v_2$ as a function of $p_\bot$ for identified particles. 
The hydrodynamic calculation results are by P. Huovinen {\it et al.}~\cite{huovinen}.}
\label{fig:v2}
\end{figure}

Figure~\ref{fig:v2} shows the $v_{2}$ as a function of $p_\bot$ for $\pi $, $K$, $p$, 
$\Lambda $ and $\Xi $ from PHENIX~\cite{phenix-v2} and STAR~\cite{star-v2} 
measurements. The angular anisotropy $v_{2}$ reveals
three salient features:
\begin{enumerate}
\item
particles exhibit hydrodynamic behavior in the
low $p_\bot$ region --- a common expansion velocity may be established from the
pressure of the system and the heavier the particle the larger the momentum
from the hydrodynamic motion leading to a decreasing ordering of $v_{2}$
from $\pi $, to $K$ and $p$ for a given $p_\bot$;
\item
 $v_{2}$ values do not
depend strongly on $p_\bot$ at the intermediate $p_\bot$ region in contrast to
the strong $p_\bot$ dependence in the yield of particles;
\item
 the saturated  
$v_{2}$ values for baryons are higher than those for mesons and there is a
distinct grouping among mesons and baryons.
\end{enumerate}

The absence of a strong $p_\bot$ dependence at the intermediate $p_\bot$
region is an intriguing phenomenon for the angular anisotropy of particle
emission.  Parton energy loss in the dense medium created in $AA$ collisions
was proposed as a possible mechanism for generating an angular anisotropy 
$v_{2}$. High energy partons are quenched inside the dense medium and lead to an
effective particle emission just from a shell area of the participating volume~\cite{muller}.
Jet quenching   scenario~\cite{Vitev} (parton energy loss scenario)
cannot explain the particle dependence in both the
nuclear modification factor and the angular anisotropy $v_{2}$ at the
intermediate $p_\bot$ region.  Within the parton energy loss scenario the
larger $v_{2}$ of baryons implies a higher energy loss than that of mesons;
the larger nuclear modification factor of baryons, however, is only
consistent with a smaller energy loss than that of mesons. 
This scenario may be important for the considerable $v_{2}$
magnitude for charged hadrons at a $p_\bot$ greater than 6 GeV/c or so
measured by STAR~\cite{star-highpt-v2}. 

Indeed, the magnitude of the measured $v_{2}$ is
significantly larger than what can be accommodated based on particle
emission from a geometrical ellipsoid source within an energy loss scenario~\cite{shuryak}.
Using a more realistic Wood--Saxon description of the colliding nuclei for the
ellipsoid source, the predicted theoretical $v_{2}$ is much smaller than that from
a hard-sphere model of the colliding nuclei, leading to a greater
discrepancy between the measurement and the theoretical expectation. 

We conclude that the magnitude and the particle dependence of $v_{2}$ at the
intermediate $p_\bot$ region cannot have a dynamical origin either 
from hydrodynamic flow or from parton energy loss alone.
Another   plausible procedure  would be to relate the $v_{2}$  in the
intermediate $p_\bot$ region to the geometrical shape of the emitting particle 
source, see figure~\ref{schem}. A
surface emission pattern from the almond shaped participant volume should
naturally lead to a saturation of $v_{2}$. A surface emission scenario is
possible if particles are produced by surface related dynamical
instabilities and the
hadronization duration is relatively short~\cite{suddenPRL}.

\subsection{Quark Flow Anisotropy}
\begin{figure}[htb]
\includegraphics[height=0.32\textheight]{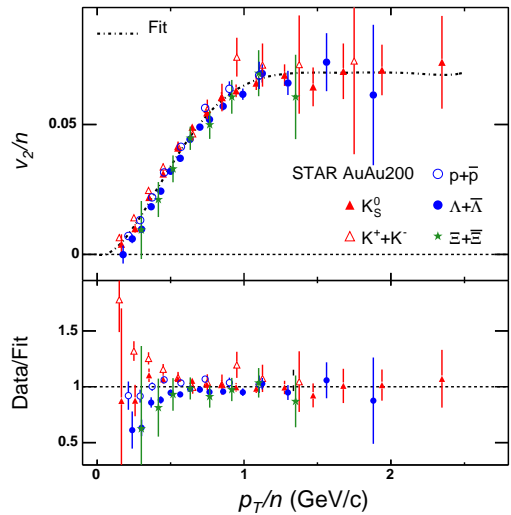} 
\caption{Azimuthal angular anisotropy $v_2/n$ as a function of $p_\bot/n$ for identified
particles where n is the number of constituent quarks. The line is a polynomial fit to the 
data points excluding the
$\pi$ data. The bottom panel shows the (multiplicative) deviation from the fit line.}
\label{scale-v2}
\end{figure}

An empirical quark number $n$ scaling of $v_2$ has  been 
noted~\cite{sorensen}. $n$ is the number of valance quarks and antiquarks 
in a hadron.  Figure~\ref{scale-v2} presents
the $v_{2}/n$ as a function of $p_\bot/n$ for   $K$, $p$, $\Lambda $
and $\Xi $ from AuAu200 collisions, where the line is a polynomial fit 
to the data points. The bottom panel shows the ratio of data points
to the fit. At the intermediate $p_\bot$ region ( $0.6<$ $p_\bot/n$ $<2$
GeV/c), all meson and baryon data points fall onto an uniform curve.
The $\pi $ data (not shown) are significantly above the $v_{2}$ of other
mesons. The large fraction of resonance decay contribution to the $\pi $
yield is known to enhance the $v_{2}$ of $\pi $ at a given $p_\bot$~\cite{Greco,Dongx}.

For  $p_\bot/n<0.6$\,GeV,   there appears  a small residual but 
systematic particle  dependent deviation from the fit curve,
which agrees best with  the $K_{\rm S}$ results. 
The residual  mass and/or strangeness and/or meson--baryon 
dependence of this deviation can have many
causes involving  microscopic transport and/or 
collective flow phenomena. Its detailed understanding could play an
important role in the demonstration of physics processes that lead to the
constituent quark level azimuthal flow at hadronization.

We believe that the particle
dependence on $v_{2}$,  which is largley explained by $n$-scaling, requires a 
$v_{2}$ distribution at the constituent
quark level. The
scaled azimuthal angular anisotropy ($v_{2}/n$) may be interpreted as the
constituent quark anisotropy just prior to the hadron formation.
 The $n$-scaling works both for strange and nonstrange quarks, indicating
that the azimuthal flow anisotropy is the same for all three `light' flavors, 
and by extension, that the 
collective quark flow is the same for the three flavors.  
The precision of the $n$ scaling  
leaves very little space for the  presence of  gluon (fragmentation) participation 
in the intermediate momentum particle production 
process. This suggests that semi-hard 
 gluons have effectively disapperad as independent partonic
degrees of freedom in the  final hadron formation.

We have shown in this section that the  hadron formation 
dynamics, at RHIC at $0.6<p_\bot/n<2$,
GeV/c is very different from the parton fragmentation
picture where the leading parton plays a dominant role in determining
properties of the final state hadron.  
The measured features for these intermediate $p_\bot$
hadrons produced at RHIC require that all quark ingredients ($n=2$ for mesons
and $n=3$ for baryons) play an approximately equal role in the
 hadron formation and that
the hadron properties are determined by the sum of contributing partons.

\section{Deconfined Bulk Matter}
\subsection{Bulk properties as function of centrality}
The global study of   soft hadron yields and spectra implies that we can  
fully characterize  the properties of bulk matter at time of hadronization
by adding up strangeness,  entropy, etc., contained in the final
 state hadrons \cite{RafActa03}. This is done particularly easily
within the statistical hadronization scheme, which  have been successfully 
applied to describe stable particle production.

As a first step, this leads  to a quantitative understanding of
 the variation of the bulk properties with centrality \cite{RafCent}.  
In such an analysis of $\sqrt{s_{\rm NN}}=200$ GeV STAR and PHENIX results,  
the contents in entropy, baryon number, strangeness, and energy grows
linearly with the participant number for  $A>20$.  This implies that the 
conversion of colliding nuclear matter into bulk parton matter is 
at RHIC  rather independent of the size of  nuclear overlap and 
complete, including all matter. In our above discussion of $v_2$, 
we have tacitly  assumed that the properties of bulk partonic
matter formed in non central collisions are not 
significantly dependent on centrality, as the present study confirms

 In most central 5\%
of the reactions, one finds that the  net baryon density per unit 
of rapidity is at mid-rapidity
$d(B-\overline B)/dy=14\pm2$, the strangeness yield $ds/dy=135\pm10$
and the entropy yield $dS/dy=4900\pm400$. The errors comprise the 
uncertainty of the data and the chemical scheme used. 
The strangeness per entropy yield 
$s/S=(2.9\pm0.3)10^{-3}$ is more than 
4 times enhanced compared to the 
AGS energy scale.  For the most peripheral 
reactions, this ratio is  $s/S=(1.9\pm0.3)10^{-3}$ which shows  
the influence of the fireball  expansion dynamics on production 
of strangeness.

\begin{figure}[htb]
\includegraphics[height=0.4\textheight]{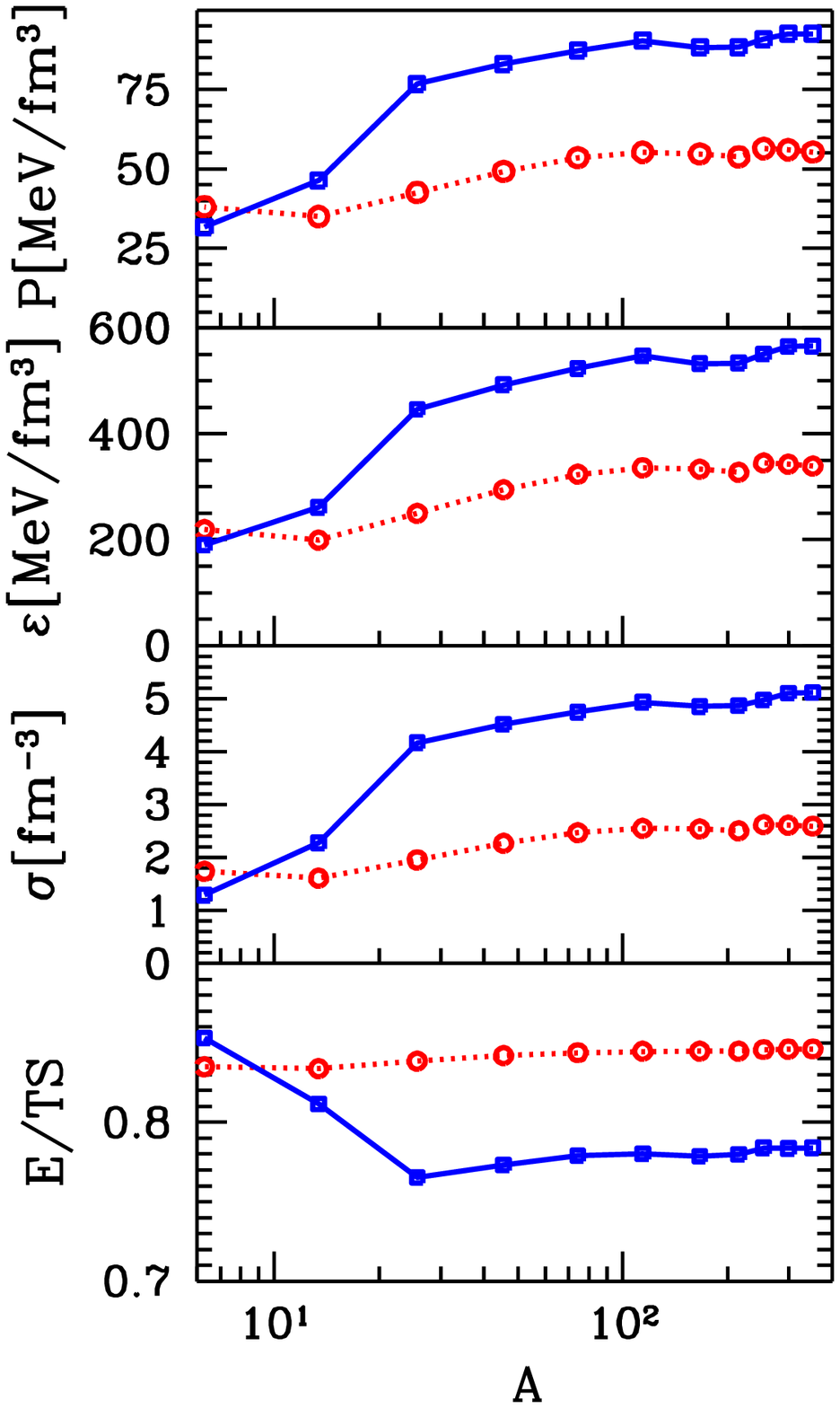} 
\includegraphics[height=0.4\textheight]{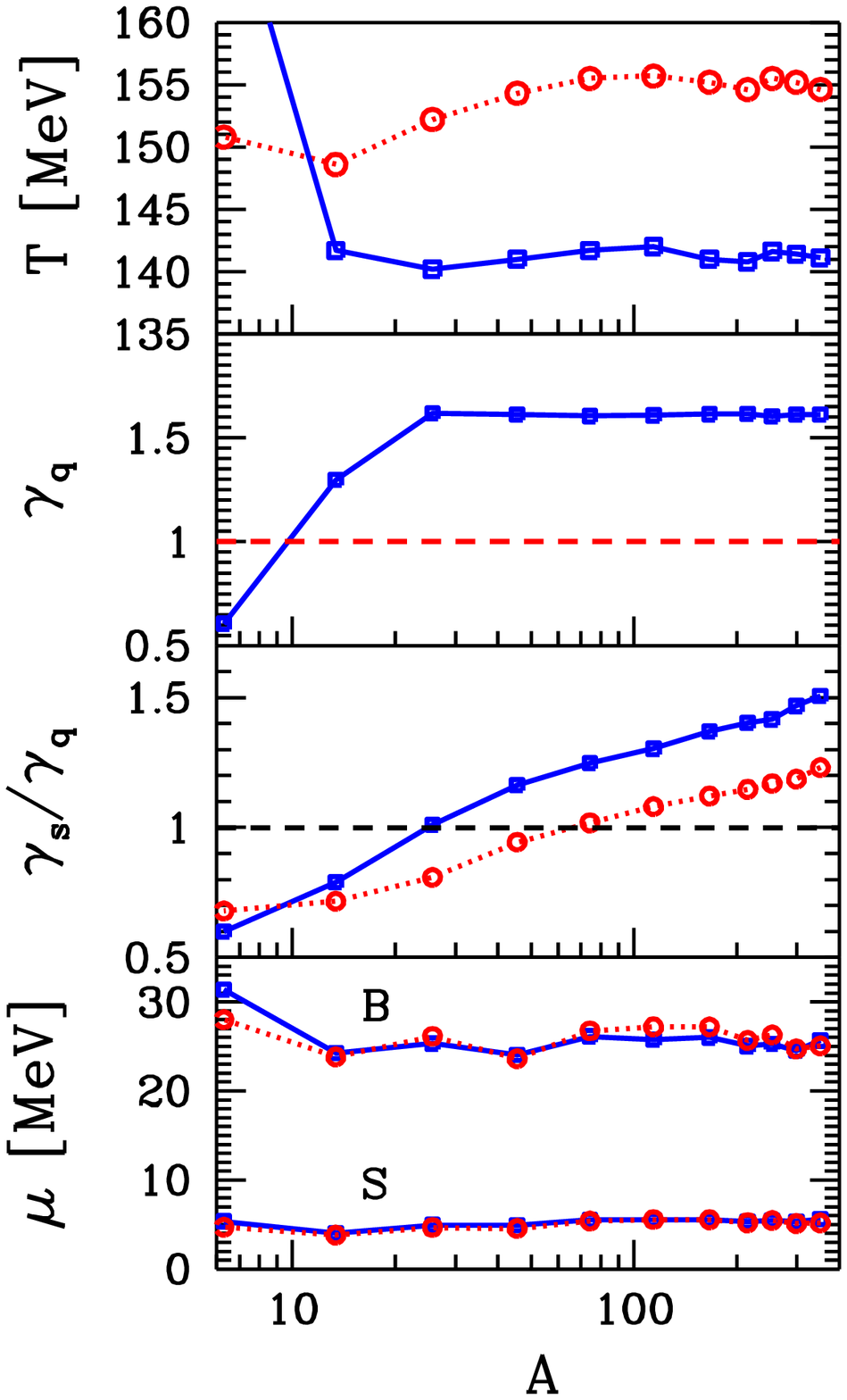} 
\caption{Left:  pressure $P$, energy density $\epsilon=E/V$, 
entropy density $S/V$ and   $E/TS$  and;
  Right: temperature $T$, light quark phase 
space occupancy $\gamma_q$,
the ratio  of strange to light quark phase 
space occupancies $\gamma_s/\gamma_q$  and the chemical potentials
($B$ for baryochemical $\mu_B$ and $S$ for strangeness  $\mu_S$) 
as a function of centrality. After Ref. \cite{RafCent}.   
 }
\label{PropCent}
\end{figure}

Given  the experimental yields of particles  at central rapidity, 
 the intensive properties  of bulk matter at 
hadronization, such as $P$ pressure, $\epsilon$ energy density, $\sigma$ 
entropy density, $E/TS=\epsilon/T \sigma$, seen in figure\,\ref{PropCent}, 
follow summing the properties of individual particle fractions.   
They  are given for the two chemical models 
in figure\,\ref{PropCent} on left. The lines
guide the eye, the actual results are the squares and circles  centered
at the mean trigger centrality of the data considered. These  results 
are   for the full chemical non-equilibrium (squares) and for 
strangeness chemical  non-equilibrium (circles) respectively. 

To obtain the  physical properties,  a fit of the statistical hadronization
model parameters had been performed and the results  are  shown on right: 
the hadronization temperature
$T$, phase space occupancies $\gamma_q, \gamma_s/\gamma_q $, and the chemical
potential $\mu_{\rm B, S}$ of baryon number and strangeness, respectively. 
We see the expected rise of strangeness occupancy yield with size (centrality) of the system,
related to the extended lifespan during which strangeness is produced \cite{Raf99,Let96}.
Otherwise there is remarkable stability of the bulk properties with centrality of the 
collision. The bulk matter created at RHIC  at $\sqrt{s_{\rm NN}}=200$~GeV, 
small or large in size and  independently 
of models of chemical (non)equilibrium has a baryochemical potential
 $\mu_{\rm B}=25\pm1$ MeV.

\subsection{The Phase Boundary}

\begin{figure}[htb]
\includegraphics[height=0.30\textheight]{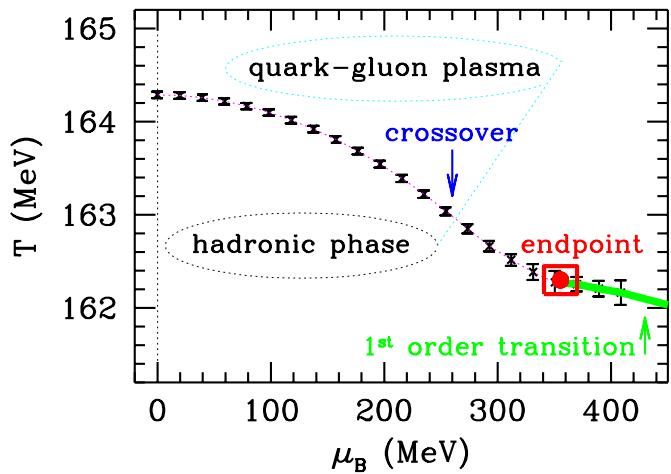} 
\caption{The phase diagram at low baryon density 
in physical units,   results of Ref. \cite{Fodor:2004nz}. 
Phase crossover ends in the   square which shows the endpoint
where 1st order phase transition sets in.
 }
\label{Phases} 
\end{figure}

The hadronization temperature $T= 140$--$155\pm8$ MeV is near 
but below  the phase transformation boundary 
for QCD with 2+1 flavors. The infinite matter static equilibrium transformation  
 is expected at  $T=164\pm4$ MeV, for more detail about the structure of the cross-over
region,  see  figure\,\ref{Phases}.  

The lower   RHIC freeze-out temperature  we reported above is 
associated with  a rapid  expansion of the RHIC fireball: the flow of 
color charge pushes the vacuum and the bulk breakup 
occurs suddenly from an over expanded 
supercooled condition. Quantitatively the supercooling  is of the required 
magnitude \cite{suddenPRL}. The sudden
hadronization  mechanism also explains why single freeze-out analysis of
particle spectra at SPS and RHIC succeed \cite{NJP,Bron01}. 
 
\section{Heavy Quarks}
\subsection{Theoretical remarks}
Heavy quarks (charm, in future bottom)  are produced mostly in initial 
parton scattering (gluon fusion) and in any case   during the initial 
primary and/or
very high temperature phase of the collision. Therefore, heavy quark
measurement can probe the initial parton flux, the dynamical evolution and the quark
energy loss in dense medium. If heavy quarks are found to participate in the
collective motion of the medium (radial or elliptic flow), this will lend
further confirmation for parton collectivity.

At RHIC, one expects that multiple heavy quark--antiquark pairs will be produced in
an individual $AA$ collision. Therefore one can expect a novel mechanism 
of $J/\Psi$ formation by recombination \cite{Thews01}. 
Then, there should be a contribution to heavy quarkonium formation
which utilizes combinations of initially uncorrelated
quark and antiquark, leading to a {\it quadratic} increase with
the total number of heavy flavor quarks in the event.  This can
occur within  the deconfined phase, considering that
recent Lattice QCD (LQCD) calculations indicate that
spectral functions of pseudo scalar and vector mesons have non-trivial shapes
at a temperature above the critical $T_{c}$~\cite{Pete}. In
particular, heavy quarkonium such as J/$\psi $ may survive at a temperature
above $1.6T_{c}$~\cite{Hatsuda,Degal}.

Charm quark transport dynamics in dense nuclear medium will provide unique
probes to the QCD properties of the medium. 
If the initial temperature is very high, $T\sim 500$ 
MeV or higher, the yield of total charm quarks can also be increased through
thermal gluon--gluon scatterings. 
Possible suppression of charm
mesons at high $p_\bot$ will test the energy loss dynamics of charm quark
propagation in a QCD medium. Theoretical calculations have predicted a reduced
medium induced energy loss for heavy quarks and the high $p_\bot$ suppression
of charmed hadrons should not be as strong as light hadrons~\cite{Kharzeev,MG}. 
For a recent review of theoretical charm quark situation see Ref.\,\cite{Thews04}.

\subsection{RHIC experimental status}
Both the STAR and PHENIX collaborations at RHIC have been
pursuing vigorous heavy quark physics programs, both in analysis of current data and in 
future detector upgrade
plans, to provide better capabilities for heavy quark measurements.

\begin{figure}[htb]
\includegraphics[height=0.35\textheight]{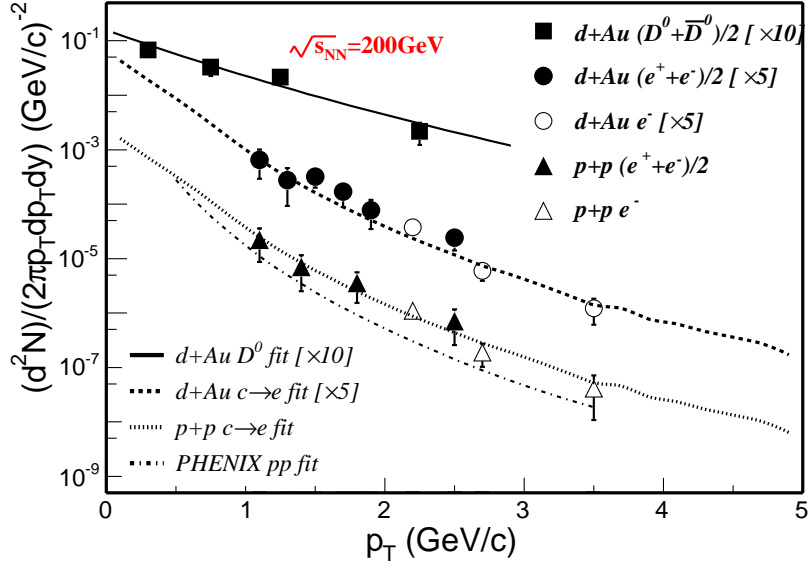} 
\caption{$p_\bot$ distribution of $D^{0}$ mesons and non-photonic 
electrons from semi-leptonic decays of charm mesons.}
\label{pt-ele}
\end{figure}

The total charm quark pair production cross section ($\sigma _{c\overline{c}}$) 
is an important constraint on the collision dynamics and the heavy quark
evolution. Both STAR and PHENIX have presented results on the 
$\sigma _{c\overline{c}}$ measurement of $pp$ collisions from charm semi-leptonic
decays. In addition, STAR has also derived an equivalent 
$\sigma _{c\overline{c}}$ for $pp$ collisions based on direct reconstruction of
hadronic decays of charm mesons from dAu collisions. Figure~\ref{pt-ele} shows the $p_\bot$-spectra of
electron and $D^{0}$ from STAR. 
The PHENIX preliminary non-photonic electron data are represented by
the fitted line with a reported measurement $\sigma _{c\overline{c}}=709\pm 85$%
(stat)$+332-281$(sys) $\mu b$~\cite{phenix-e}. 
STAR has measured $\sigma _{c\overline{c}}=1300\pm
200\pm 400$ $\mu b$ and a mean transverse momentum for $D^{0}$ of 
$\langle p_\bot\rangle=1.32\pm 0.08$ GeV/c from direct $D^{0}$
 reconstruction~\cite{star-e}.  A next-to-leading order pQCD
calculation of the charm quark production cross section~\cite{vogt} 
has yielded $\sigma _{c\overline{c}}=300$ --- 450 $\mu b$,
significantly below the STAR measurement and at the lower end of the PHENIX
range of uncertainty.

Several comments on the cross section measurements are in order. When
measuring the charm cross section through semi-leptonic decay, 
the quality of the electron data for $p_\bot\lessim 1$ GeV/c  
is significantly deteriorated because of a large combinatorial background.
An electron of $p_\bot$ $\sim 1$ GeV/c typically comes from the decay of a 
$D$ meson with $p_\bot$ $\sim 2$ GeV/c, which is significantly beyond
the average $p_\bot$ of $1.32\pm 0.08$ GeV/c reported by STAR. Therefore, one
has to extrapolate to the low $p_\bot$ region by over a factor of two to
obtain the total charm cross section. Such an extrapolation is often model
dependent and has a large uncertainty. The semi-leptonic decay branching
ratios for $D^{0}$, $D^{\ast }$, $D^{\pm }$ and $D_{S}$ are different. The
electron yield from decays of these $D$ mesons depends on the relative yield
which is another important contribution to the uncertainties of the charm
production cross section derived from electron measurement. The direct
reconstruction of the $D$ decay kinematics provides a broad coverage of 
$p_\bot$ and does not suffer from the same uncertainties as the leptonic decay
electron measurement. However, present STAR measurements of $D$ meson yields using
event-mixing methods from TPC tracks suffer from limited statistics. A future
vertex detector upgrade capable of measuring the $D$ decay vertex
displacement is essential for both STAR and PHENIX heavy flavor physics
programs.

\begin{figure}[htb]
\includegraphics[height=0.35\textheight]{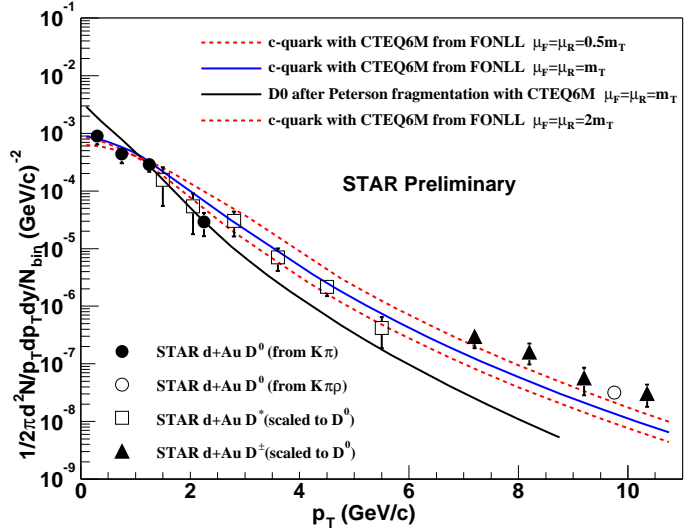} 
\caption{STAR preliminary measurement of $p_\bot$ distribution of $D$ mesons from dAu collisions normalized 
by the number of binary collisions. The shape of the $p_\bot$ distribution is compared to pQCD FONLL calculations.}
\label{D-pt}
\end{figure}

Figure~\ref{D-pt} presents the STAR preliminary transverse momentum spectrum of $D$
mesons from dAu collisions normalized by the number of binary collisions,
where the $p_\bot$ shapes of $D^{\ast }$ and $D^{\pm }$ are assumed to be the
same as that of $D^{0}$~\cite{atai}. The shape of the $p_\bot$ distribution coincides
with the bare charm quark $p_\bot$ distribution from the Fixed-Order-Next-Leading-Log (FONLL)
pQCD calculation from M. Cacciari {\it et al.}~\cite{fonll}. If a fragmentation function such as the
Lund fragmentation scheme~\cite{pythia} or the Peterson function~\cite{peterson} is introduced
for $D$ meson production, the resulting $p_\bot$ distribution will be
significantly below the measurement at the high $p_\bot$ region. This
observation raises an outstanding question regarding the $p_\bot$
distribution and the formation mechanism of $D$ mesons in hadro-production.
Recently,
a $k_\bot$ factorization scheme has been found to significantly change the $D$ meson $p_T$
distribution from nuclear collisions as well~\cite{Tuchin}.

The fact that the $D$ meson $p_\bot$ distribution can be better described by
the $p_\bot$ of bare charm quarks from pQCD calculations has been observed in
previous fixed target experiments~\cite{FONLL-1}. With a fragmentation function such as
the Peterson function the $D$ meson $p_\bot$ distribution is too steep to
explain the measured $p_\bot$ distribution. In order to match the calculation
with the data one has to introduce a $k_\bot$ kick to the parton
distribution. The scale of $\langle k_\bot^{2}\rangle $ is on the order of 1 (GeV/c)$^{2}$,
much larger than the typical $\Lambda _{QCD}$ scale for strong
interaction. Furthermore, the Feynman $x_{F}$ of the $D$ meson distribution
was also found to coincide with the bare charm quark $x_{F}$ distribution~\cite{FONLL-1}. 
The fragmentation function will have a large impact on the $x_{F}$
distribution from bare charm quark to $D$ meson, which cannot be negated by
introducing any $k_{L}$ longitudinal boost of reasonable scale as in
the case for $k_\bot$ kick in the transverse momentum direction.

The transverse momentum distribution of particles produced at RHIC energies
are considerably flatter than those at lower energies. The $k_\bot$ kick
scheme does not change the shape of the $p_\bot$ distribution significantly.
The STAR measurement of the $D$ meson $p_\bot$ distribution suggests either that
the charm quark fragmentation may be close to a delta function or that a charm
formation mechanism such as the recombination model may be important in
hadro-production. The recombination model takes a charm quark and combines
it with a light quark, presumably of low $p_\bot$, to form a charmed meson.
Therefore, the $p_\bot$ of the meson is not significantly different from that
of the bare charm quark. In that way, measurements of charmed meson production 
provide unique probes for the hadron formation dynamics
and for the transport dynamics of heavy quarks in the dense nuclear medium 
produced at RHIC energies.

\begin{figure}[htb]
\includegraphics[height=0.35\textheight]{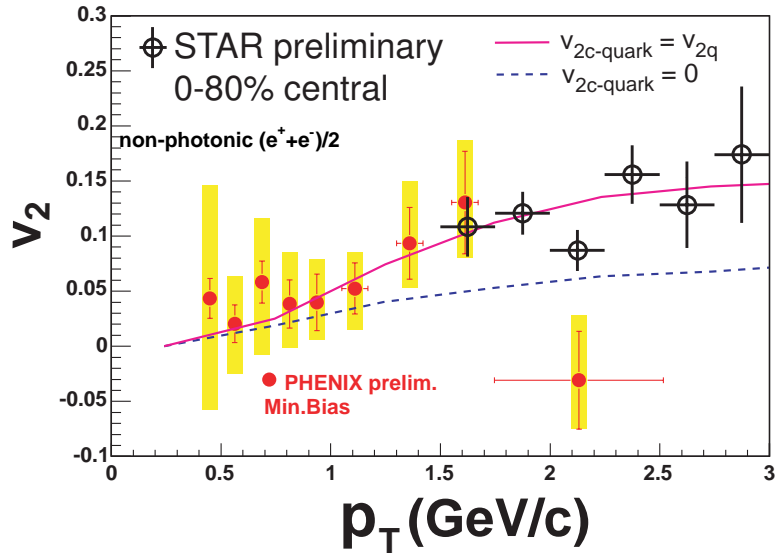} 
\caption{Preliminary measurement of non-photonic electron $v_2$ as a function 
of $p_T$ from PHENIX and STAR. The curves are from a calculation by V. Greco {\it et al.}~\cite{Rapp}.}
\label{ele-v2}
\end{figure}

Observation
of heavy quark hydrodynamic flow would indicate that heavy quarks, once
created in the initial state, must have participated in the partonic
hydrodynamic evolution over a sufficiently long period of time to reach a
substantial flow magnitude. This would be a unique probe for the early stage
of a partonic phase~\cite{Rapp}. Figure~\ref{ele-v2} shows preliminary
STAR~\cite{star-e-v2} and PHENIX~\cite{phenix-e-v2}  measurements of the $v_{2}$ for 
electrons from charm leptonic
decays, which have been demonstrated to be closely correlated with charmed
meson $v_{2}$~\cite{Dongx}.

\section{Summary}
Among the most important aspect of many  results presented is that the
gluon degrees of freedom are not explicitly manifested in the hadron formation.
Quarks are the dominant  degrees of freedom
at the boundary of quark--hadron transition.  
It is a critical step to firmly connect this mainly
 experimental insights on the
properties of the quark matter at the boundary of hadronization with the
LQCD calculations. The disappearance of the gluon degrees of freedom from
the initial state and the emergence of constituent quarks at
hadronization are some of the critical conceptual questions to be addressed
in the imminent future \cite{Maas}.

 Strange (anti) baryon production enhancement at RHIC 
as function of centrality, energy   and $p_\bot$ is significant.
For   $p_\bot\lessim 5$  GeV/c,  relative baryon to meson yields
 have shown distinct features that are
drastically different from the fragmentation processes in elementary collisions.
A novel  feature of meson and baryon dependence has been observed in the
nuclear modification factor and the angular anisotropy $v_{2}$ of $\pi $, 
$K^{\pm }$, $K_{S}$, $K^{\ast }$, $p$, $\Lambda $ and $\Xi $ particles at
the intermediate $p_\bot$ of 2--5 GeV/c.  
  A constituent quark number scaling has been observed
for the $v_{2}$ measurement.  These experimental measurements suggest that 
 quarks have developed a collective $v_{2}$ 
as a function of $p_\bot$; and the hadron formation at the intermediate 
$p_\bot$ is likely through multi-parton dynamics such as 
recombination or coalescence process.

This physical picture emerging from the experimental measurements complements
the Lattice QCD results. Spectral function calculations have indicated that
hadrons, particularly heavy quarkonium, do not melt completely at critical
temperature. It appears plausible that the constituent quark degrees of
freedom or even hadron-like 
quasi-particles play a dominant role at the hadronization of bulk
partonic matter though further confirmation of the picture from LQCD is
needed. We note the  related theoretical models~\cite{brown},  
 which invoke the notion
of quasi-hadrons to describe properties of the dense matter.
 
We have also described how heavy (charm) 
quark production and its transport dynamics in dense nuclear medium
probe QCD properties of the matter. The charm quark flow measurement will
provide a significant insight on recombination or coalescence hadronization
mechanism and partonic collectivity of the dense matter. Charmonium 
enhancement arises from a dense charm rich state. Future detector
upgrades from STAR and PHENIX will greatly enhance their heavy quark
measurement capabilities at RHIC.

 Despite intriguing experimental observations of hadronization from a
deconfined bulk partonic matter, a smoking colt signatures for the quark--hadron phase
transition remain elusive. On the other hand, the totality of experimental 
results can   be understood invoking deconfinement.

\begin{theacknowledgments}
We thank An Tai, Hui Long, Paul Sorensen, Xin Dong, Frank Laue, Zhangbu Xu, Nu Xu, 
Charles Whitten Jr., Jean Letessier, Giorgio Torrieri, Robert Thews 
 for many stimulating discussions  on physics topics in this article.
 We thank Nora  Brambilla,  
   Giovanni Prosperi, and the local organizers  of the "VI Quark Confinement
and the Hadron Spectrum" conference, for their very kind hospitality in
Sardinia.

\end{theacknowledgments}

\end{document}